%% file: main.tex
\documentclass[10pt,twocolumn,letterpaper]{article}

\usepackage[pagenumbers]{cvpr} % To force page numbers, e.g. for an arXiv version


\definecolor{cvprblue}{rgb}{0.21,0.49,0.74}
\usepackage[pagebackref,breaklinks,colorlinks,allcolors=cvprblue]{hyperref}
\usepackage{hyperref}
\usepackage{url}
\usepackage{listings}
\usepackage{longtable,booktabs}
\usepackage{float}
\usepackage{graphicx}
\usepackage{epstopdf}
\usepackage{multirow}
\usepackage{amssymb}% http://ctan.org/pkg/amssymb
\usepackage{pifont}% http://ctan.org/pkg/pifont
\newcommand{\cmark}{\ding{51}}%
\newcommand{\xmark}{\ding{55}}%
\usepackage{tikz}

\def\halfcheckmark{\tikz\draw[scale=0.4,fill=black](0,.35) -- (.25,0) -- (1,.7) -- (.25,.15) -- cycle (0.75,0.2) -- (0.77,0.2)  -- (0.6,0.7) -- cycle;}

\lstdefinestyle{myverbatim}{
    basicstyle=\ttfamily\footnotesize,
    backgroundcolor=\color{white},
    breaklines=true,
    breakatwhitespace=true
}

\newsavebox{\measurebox}

\UseRawInputEncoding

\def\paperID{9280} % *** Enter the Paper ID here
\def\confName{CVPR}
\def\confYear{2025}

\title{Audio-Agent: Leveraging LLMs for Audio  Generation, Editing and Composition}

\author{
Zixuan Wang$^{1}$, 
Chi-Keung Tang$^{1}$,
Yu-Wing Tai$^2$ \\
$^1$HKUST  \quad $^2$Dartmouth College \\
}

\begin{document}
\maketitle
\input{sec/0_abstract}    
\input{sec/1_intro}

\input{sec/2_related_work}

\input{sec/3_method}
\input{sec/4_experiments}
\input{sec/5_conclusion}
{
    \small
    \bibliographystyle{ieeenat_fullname}
    \bibliography{main}
}

\input{sec/X_suppl}

\end{document}

%% file: sec/0_abstract.tex
\begin{abstract}
We introduce Audio-Agent, a multimodal framework for audio generation, editing and composition based on text or video inputs. Conventional approaches for text-to-audio (TTA) tasks often make single-pass inferences from text descriptions. While straightforward, this design struggles to produce high-quality audio when given complex text conditions. In our method, we utilize a pre-trained TTA diffusion network as the audio generation agent to work in tandem with GPT-4, which decomposes the text condition into atomic, specific instructions and calls the agent for audio generation. In doing so, Audio-Agent can generate high-quality audio that is closely aligned with the provided text or video exhibiting complex and multiple events, while supporting variable-length and variable-volume generation. For video-to-audio (VTA) tasks,  most existing methods require training a timestamp detector to synchronize video events with the generated audio, a process that can be tedious and time-consuming. Instead, we propose a simpler approach by fine-tuning a pre-trained Large Language Model (LLM), e.g., Gemma2-2B-it, to obtain both semantic and temporal conditions that bridge the video and audio modality. Consequently, our framework contributes a comprehensive solution for both TTA and VTA tasks without substantial computational overhead in training.
\end{abstract}

%% file: sec/1_intro.tex
\section{Introduction}
Multimodal deep generative models have gained increasing attention these years. These models are trained to perform tasks based on different kinds of input called modalities, mimicking how humans make decisions from different sensory inputs such as vision and smell~\cite{suzuki2022survey}. Compared to other generation tasks such as image generation or contextual understanding, audio generation is less intuitive as it is harder to precisely measure the generated sound quality using human ears. Moreover, previous works mainly focus on generating music-related audio, which is more structured compared to naturally occurring audio~\cite{copet2024simple,melechovsky2023mustango}. Some recent works have focused on generating visually guided open-domain audio clips~\cite{chen2020generating,zhou2018visual}. 

\begin{figure*}
    \centering
    \includegraphics[height=0.3\textheight]{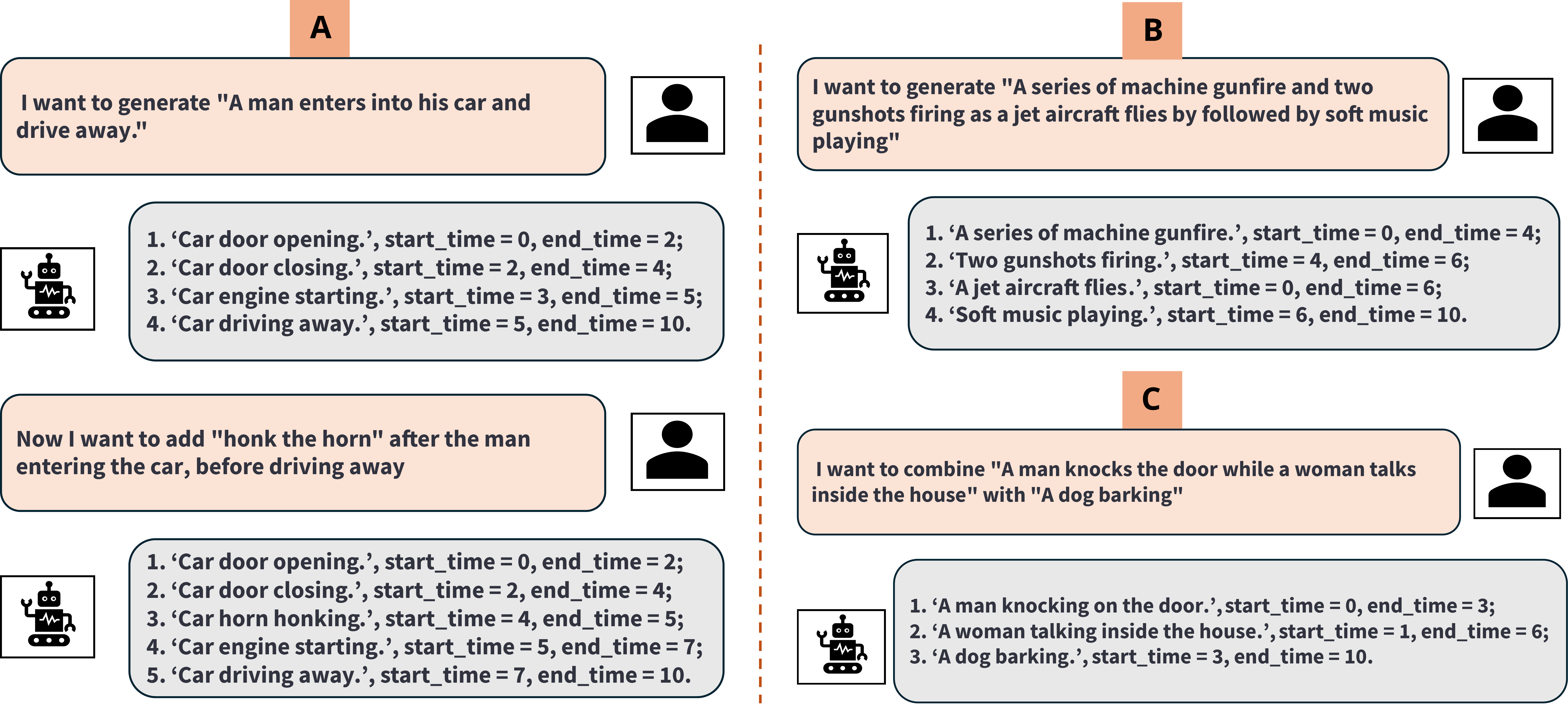}
    \vspace{-0.1in}
    \caption{Example showing Audio-Agent's ability to generate, compose and edit multiple audio descriptions together. (A): Multi-turn editing; (B): Generation based on long description; (C): Multiple audio descriptions composition.}
    \label{fig.demo_case}
    \vspace{-0.15in}
\end{figure*}

Recent research on audio generation has mainly focused on text-to-audio generation (TTA) and video-to-audio generation (VTA). For TTA~\cite{xue2024auffusion,kreuk2022audiogen}, current datasets lack high-quality text-audio pairs. Existing datasets such as AudioCaps~\cite{kim2019audiocaps} or Clotho~\cite{drossos2020clotho} usually contain multiple event descriptions mixed into a single sentence without fine-grained details and object bindings. This complicates training, particularly when handling long continuous signals with complex text conditions~\cite{huang2023make}. We define complex text conditions as long event descriptions containing a series of events without explicitly describing the sound, such as ``A man enters his car and drives away". While previously not fully studied, this type of condition is more realistic, as it does not require any detailed specification in terms of the characteristics of the audio result, offering more flexibility to users and producers in
movie dubbing and musical composition. 

The VTA task, or conditional Foley generation, has remained relatively unexplored until recently~\cite{wang2024c3llm,zhang2024foleycrafter}. One main challenge is that video clips typically contain excessive visual information not always relevant to audio generation. Moreover, synchronization is hard between video and audio output, with recent solutions such as temporal masking~\cite{xie2024sonicvisionlm} showing still much room for improvement to handle %inadequate for 
complex scenarios. Due to efficiency considerations, current methods often encode video features by extracting a few random frames~\cite{xie2024sonicvisionlm,dong2023clipsonic}, which hinders learning temporal information. Bridging the modality gap~\cite{liang2022mind} between video and audio thus becomes the key to solving the problem. 

While achieving state-of-the-art results, conventional approaches perform inference in a single pass based on a given text description. This approach struggles to produce high-quality audio when given complex or lengthy text conditions as input. In this paper, we introduce {\bf Audio-Agent}, which employs GPT-4 to delineate intricate user inputs using GPT-4 into multiple generation steps. Each step includes a description along with start and end times and optionally volume control to effectively guide the audio generation process. Our general framework integrates two key tasks: Text-to-Audio (TTA) and Video-to-Audio (VTA). 

First, we leverage a pre-trained TTA diffusion model, Auffusion~\cite{xue2024auffusion} with essential adaptations, serving as the backbone for our generation process. In the TTA task, Auffusion focuses solely on generating simple, atomic text inputs. Our framework supports audio generation, editing, and composition, as illustrated in Figure~\ref{fig.demo_case}. 

Second, for the VTA task, we recognize that models such as GPT-4 and other large language models lack sufficient temporal understanding on video clips. To address this problem, we employ moderate fine-tuning to align the two modalities. We utilize the smaller Gemma2-2B-it model, which has 2 billion parameters, and fine-tune an adapter and a projection layer to convert visual inputs into semantic tokens. We then implement cross-attention guidance between the diffusion layers of Auffusion. This approach eliminates the need for additional training on a temporal detector, as the semantic tokens inherently contain time-aligned information.
Our contributions are summarized as:
\begin{enumerate}
\item we propose Audio-Agent, which utilizes a pre-trained diffusion model as a generation agent, for both TTA and VTA tasks; 
\item for TTA, Audio-Agent can handle complex text input, which is decomposed into simpler and atomic generation conditions on which the diffusion model can effectively make inference; 
\item for VTA, we fine-tune an open-source LLM (Gemma2-2B-it) to bridge the modality gap between video and audio modalities to align the underlying %providing both 
semantic and temporal information. 
\end{enumerate}
Through extensive evaluation, our work demonstrates on-par results compared to the state-of-the-art task-specific models trained from scratch, while capable of producing high-quality audio given long and complex textual input. We hope our work can motivate more relevant works on multi-event long-condition TTA generation, which to our knowledge has not yet been adequartely explored despite its high potential in various content generations where high-quality audio is essential.

%% file: sec/2_related_work.tex
\section{Related Work}
\noindent\textbf{LLM-based Agent Method} 
Recent progress in large language models has enabled relevant research on making LLM a brain or controller for the agent on performing various tasks, such as robot task planning and execution~\cite{driess2023palm} or software development~\cite{rawles2024androidinthewild,yang2023appagent}. LLM demonstrates the capacity of zero-shot or few-shot generalization, making task transfer possible without significant change of its parameters~\cite{xi2023rise}. In our work, we harness the action-planning ability of LLM. Upon receiving the text condition from the user, LLM generates a plan with detailed steps on how to call the diffusion model which serves as a generation agent. By dividing the task into simpler sub-tasks, we can ensure the generation quality with fine-grained event control for TTA generation.

\begin{table*}
\caption{Comparison of functionalities between recent audio generation framework. For AudioLDM2 and Auffusion half check marks are assigned because the corresponding model was trained only on 10 seconds of audio clips. In theory, it also supports long audio generation, but the quality is not assured, see Figure~\ref{fig.tta-example}.}
\vspace{-0.1in}
\label{tab.comparison_functionality}
\centering
\resizebox{0.6\linewidth}{!}{%

\begin{tabular}{lc|cccc}
\toprule[1.5pt]
\multirow{2}{*}{Method} & \multirow{2}{*}{VTA generation} & \multicolumn{3}{c}{TTA generation}  \\ 
& & Multi-turn editing & Composition & Long complex  generation  \\
\midrule[1.5pt]
Diff-Foley           & \cmark                 & \xmark & \xmark & \xmark                  \\
FoleyCrafter       & \cmark                 & \xmark & \xmark & \xmark                  \\
AudioLDM2       & \xmark                 & \xmark & \xmark & \halfcheckmark                \\
Auffusion          & \xmark                 & \xmark & \xmark & \halfcheckmark                 \\
Make-An-Audio 2          & \xmark                 & \xmark & \xmark & \cmark                 \\
AudioGPT          & \xmark                 & \cmark & \cmark & \cmark                 \\
WavJourney          & \xmark                 & \cmark & \cmark & \cmark                 \\
WavCraft          & \xmark                 & \cmark & \cmark & \cmark                 \\
Ours        & \cmark                 & \cmark & \cmark & \cmark                 \\
\bottomrule[1.5pt]
\end{tabular}
}
\vspace{-0.15in}
\end{table*}

\vspace{2mm}

\noindent\textbf{Diffusion-based Audio Generation} 
AudioLDM~\cite{liu2023audioldm} is among the pioneering works that introduce the latent diffusion method to audio generation. Subsequent works such as Tango~\cite{ghosal2023text} Auffusion~\cite{xue2024auffusion}, and AudioLDM 2~\cite{liu2024audioldm} use pre-trained LLM such as Flan-T5 for text encoding, which has been widely adopted. We notice that this method can be seamlessly adapted to VTA tasks when a similarly effective way can be derived for utilizing LLM in encoding visual content. For the TTA task, we choose Auffusion as our generation agent due to its outstanding performance on fine-grained alignment between text and audio. 
\vspace{2mm}

\noindent\textbf{Coarse-to-fine Audio Generation} 
Current works such as AudioLM~\cite{borsos2023audiolm}, VALL-E~\cite{wang2023neural} and MusicLM~\cite{agostinelli2023musiclm} use multiple codebooks and Residual Vector Quantization (RVQ)~\cite{defossez2022high} to create diverse audio representations. In AudioLM, the model first predicts semantic tokens that capture crucial information for overall audio quality, such as rhythm and intonation, while subsequent layers add details to enhance the richness of the generated sound. However, these discrete designs suffer from generation quality compared to their continuous-valued counterparts. Moreover, the model has to perform prediction over multi-layers, which inevitably increases computational demands for both training and inference~\cite{meng2024autoregressive}. 

%In our case for the VTA task, we fine-tune an LLM to predict an intermediate discrete representation as semantic tokens using a language modeling approach. The discrete semantic tokens then serve as a condition for the diffusion model to generate continuous predictions. In this way, our method simplifies the generation procedure while maintaining the advantages of audio generation using the language modeling approach.

\vspace{2mm}
\noindent\textbf{TTA and VTA}
Make-an-audio 2~\cite{huang2023make} is one of the first methods capable of taking complex textual descriptions for audio generation. However, this method primarily focuses on the ordering of the generated audio, lacking the ability to further edit the generated audio through multi-turn conversations. Furthermore,~\cite{huang2023make} relies on extensive training with LLM-based data augmentation and is limited to TTA generation.
Subsequent works, such as AudioGPT~\cite{huang2024audiogpt}, WavJourney~\cite{liu2024wavjourney}, and WavCraft~\cite{liang2024wavcraft}, introduce specialized audio generators for compositional audio. However, these models still rely on off-the-shelf generators and do not fully leverage GPT-4's potential. They are also limited to text-based descriptions, and the generated audio often fails to meet user-specified length requirements, as the complexity of GPT-4's generation process exceeds their capabilities.

%Subsequent works, such as AudioGPT~\cite{huang2024audiogpt}, WavJourney~\cite{liu2024wavjourney} and WavCraft~\cite{liang2024wavcraft}, introduce the use of specialized audio generation experts for compositional audio creation. Despite these advances, these models still do not incorporate visual inputs in their design, focusing mainly on text-based descriptions for generating audio. The generated audio length often does not comply with the user's requirement, as the generation script that GPT-4 aims to produce is too sophisticated for these methods to handle.

For VTA, recent works such as FoleyGen~\cite{mei2024foleygen} and V2meow~\cite{su2024v2meow} employ transformers to model the audio generation process. However, as noted in~\cite{li2024autoregressive}, this autoregressive design can lead to suboptimal performance compared to continuous generation methods. In particular, errors tend to accumulate during inference, which can degrade the model's overall performance.

%% file: sec/3_method.tex
\section{Method}
\begin{figure*}
    \centering
    \includegraphics[height=0.32\textheight]{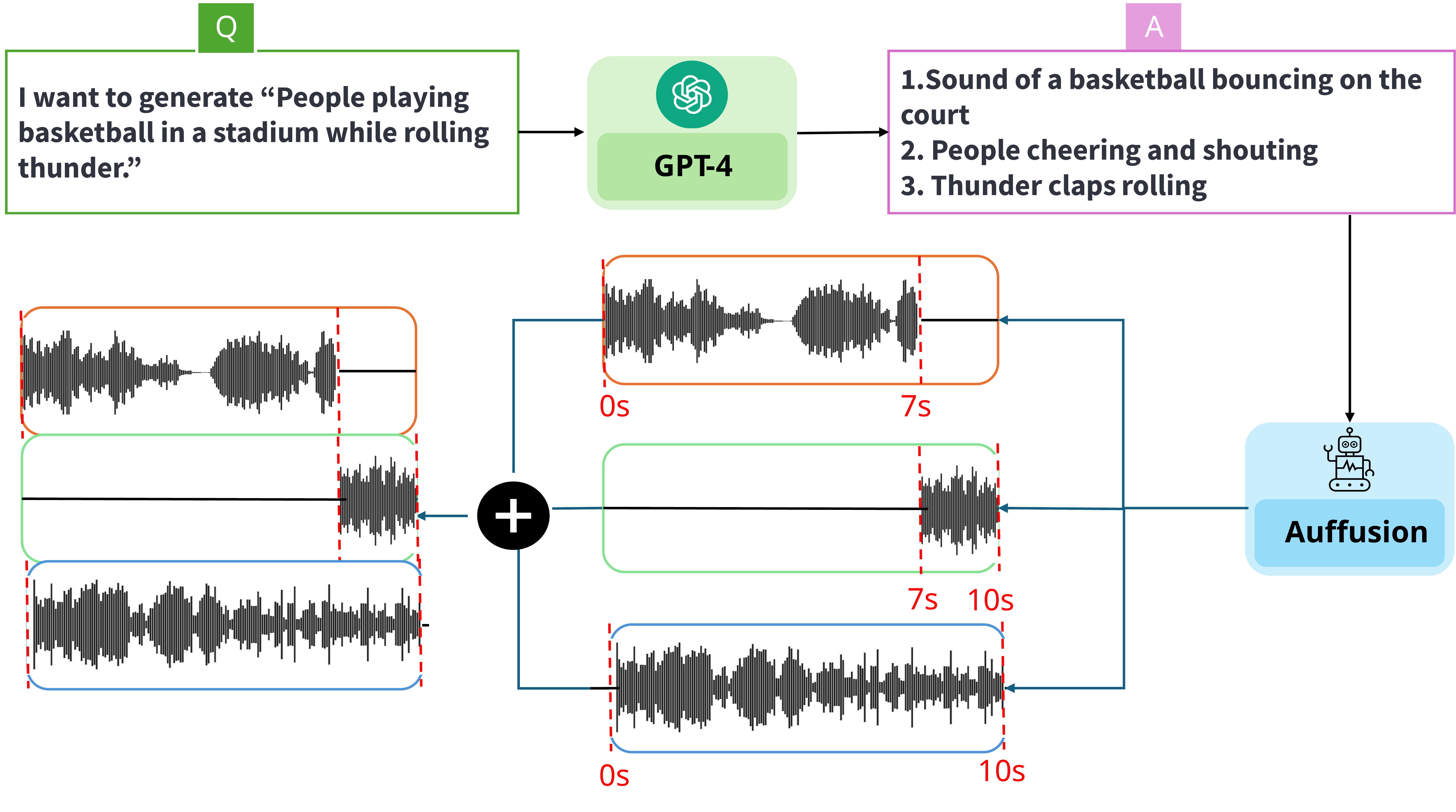}
    \vspace{-0.15in}
    \caption{Overview of the TTA part. We use GPT-4 to convert a complex audio generation process into multiple generation steps and combine inference results.}
    \label{fig.arch-tta}
    \vspace{-0.15in}
\end{figure*}

\begin{figure*}
    \centering
    \includegraphics[height=0.32\textheight]{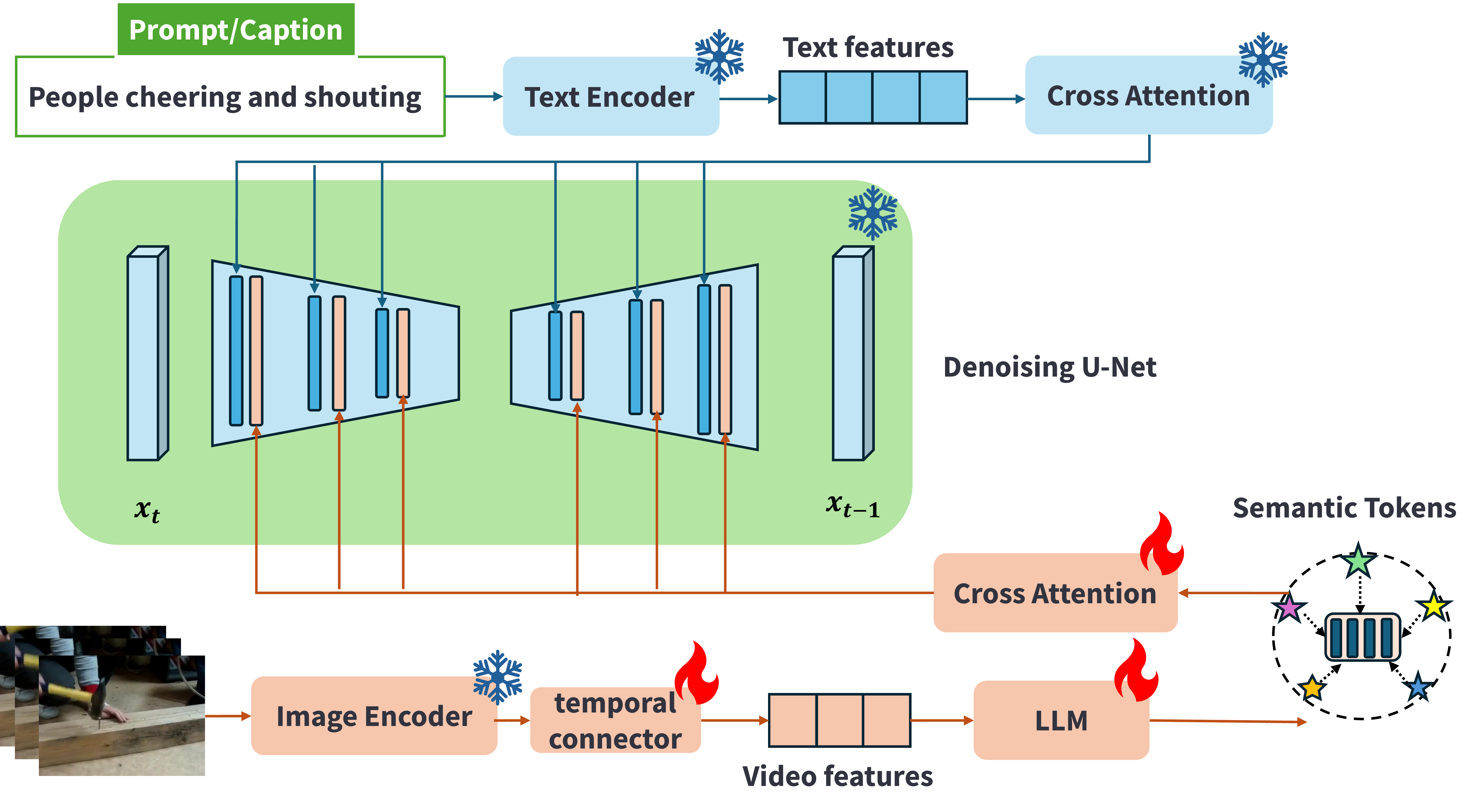}
    \vspace{-0.1in}
    \caption{Overview of the generation backbone. We build on top of the pre-trained Auffusion model for both TTA and VTA generation.}
    \label{fig.arch-backbone}
    \vspace{-0.1in}
\end{figure*}

Audio-Agent comprises three major components: 1) GPT-4 as a brain for action planning;  2) a lightweight LLM to convert video modality into semantic tokens; and 3) a pre-trained TTA diffusion model as the generation backbone. Our model structure is illustrated in Figure~\ref{fig.arch-tta} and Figure~\ref{fig.arch-backbone}.

\subsection{Preliminaries}

\noindent\textbf{Audio Latent Diffusion Model} 
Recent research adapted the successful latent diffusion models from the image domain to the audio domain. A typical audio latent diffusion model such as Auffussion first converts the audio wave into mel spectrogram, followed by  VAE encoding into the relevant latent space. Inference is the reverse process, where the predicted latent is decoded by VAE and then converted back from mel spectrogram into audio wave through a vocoder such as HiFi-GAN~\cite{kong2020hifi}. The latent diffusion process can be regarded as the same as the standard latent diffusion model on image generation~\cite{rombach2022high}. 

\vspace{2mm}

\noindent\textbf{Semantic token} AudioLM~\cite{borsos2023audiolm} was among the first to propose a two-stage method for speech synthesis. In their method, the semantic tokens are derived from representations produced by an intermediate layer of w2v-BERT~\cite{chung2021w2v}. We choose an open-sourced HuBERT~\cite{hsu2021hubert} model to produce the semantic representation, since HuBERT can model long-term temporal structure in a generative framework. Codec models such as w2v-BERT or HuBERT are originally pre-trained on speech-heavy datasets, still they exhibit the potential to be extended in other fields like music generation, as verified in~\cite{agostinelli2023musiclm}. Although only the smallest HuBERT model has its quantizer released and open-sourced, we found that the released small model is already sufficient to assist the diffusion model in generating high-quality and temporally aligned predictions.

\subsection{GPT-4 as an action planner for TTA task}\label{GPT}
Given a long, complex text condition, we ask GPT-4 to decompose the description into simple and atomic generation steps. While GPT-4 has freedom to decide the number of steps to generate, we restrict GPT-4 to keep the minimum number of necessary generation steps. This step instruction produces a good balance avoiding either extreme of being too abstract or too specific with unnecessary details. We also inform  GPT-4 that the user may revise the text requirement in subsequent conversations, so that our framework can perform multi-turn conversational generation. 

The output of GPT-4 consists of a JSON file, which contains a series of function calls to the agent model with the text description provided. In addition, to support variable length generation and multi-event generation, GPT-4 also provides the start time and end time for each call which can overlap with each other.  After obtaining the generation result for each step, we add the pertinent waveforms together based on their time range. During our experiments, we found that sometimes if the user provides compositional textual descriptions, such as combining lightning and people shouting events together, the overlapping region of the final audio may experience volume explosion. Thus, in one of our design variations, we further ask GPT-4 to control the audio volume by specifying the volume for each generation call in dB, following LUFS standard. See the supplementary material for prompt examples.

\subsection{Audio Tokenizer and Video Tokenizer}
Following~\cite{kharitonov2021text}, we utilize the 9th layer of the HuBERT-Base model to derive the semantic tokens. The quantizer of HuBERT-Base contains 500 centroids. During the training process of the VTA task, we aim to fine-tune the LLM to predict the correct semantic tokens indices from the video input. To obtain the ground truth semantic tokens indices, we first extract the ground truth audio clips from the video. We then use HuBERT acting as an audio tokenizer that applies the K-mean clustering algorithm and converts the audio into the target discrete semantic tokens indices, where each index has a value ranging from 0 to 499 to represent the respective centroids. During inference, the semantic token is constructed by fetching the corresponding centroid. HuBERT-Base has a frame rate of 50Hz, thus a 10-second audio will produce 500 semantic tokens. The HuBERT model and LLM act as a bridge that connects the video modality and audio modality. Compared with previous methods that directly use a projection layer for alignment, the semantic token serves as an intermediate representation between visual and audio modality, facilitating the conditional audio generation process.

To efficiently capture both visual and temporal information while compressing the video data, we employ CLIP as a frame-wise feature tokenizer. CLIP is compatible with any frame sampling strategies, enabling a flexible frame-to-video feature aggregation scheme as noted in~\cite{cheng2024videollama}. We pool the information within each frame to reduce the sequence size, resulting in a vector $f^r$ of size $N \times D$, where $N$ is the number of frames and $D$ is the CLIP hidden size. We set the frame rate to 21.5 Hz and use CLIP ViT-L/14 by default. 

Inter-frame information is crucial for the model to achieve temporal alignment. Previous methods~\cite{iashin2021taming,du2023conditional} require extracting both RGB and optical flow information within and across frames. In our design, we add a temporal connector after obtaining frame-wise features. The temporal connector consists of a 1D convolution block and a projection layer. The convolution block aggregates the inter-frame features together while preserving the temporal order. The projection layer projects the features into LLM's embedding space. 

\subsection{LLM for semantic token generation on VTA task}
Semantic tokens allow us to represent continuous audio information in discrete semantic form. We denote the continuous audio ground truth extracted from the video input as $a \in \mathbb{R}^{C \times L}$, where $C$ is the number of channels and $L$ is the time of the audio clip times sample rate. The HuBERT audio tokenizer applies the K-means algorithm to convert the representation into discrete indices. Specifically, we obtain the indices $s \in \{0,...,499\}^{N}$ from the audio by comparing it with the encoded audio with centroids, and $N$ is the sequence length. 

During training and inference, we feed the model with encoded video embedding and caption, together with the instruction prompt. To better differentiate the video input with text condition and instruction, we wrap the encoded video feature with special tokens as modality indicators. Specifically, we wrap the video caption with $\langle\textit{Caption}\rangle$, $\langle\textit{/Caption}\rangle$ indicators and video embedding in an embedded sequence of $\langle\textit{Video}\rangle$, $\langle\textit{/Video}\rangle$ indicators. In doing so, we avoid the possibility of confusing the LLM with different kinds of information. See supplementary material.   

To jointly model different modalities in a unified model, we further extended the LLM's inherent text vocabulary $V_t = \{v_i\}_{i=1}^{N_t}$with acoustic vocabulary $V_a = \{v_j\}_{j=1}^{N_a}$. The acoustic vocabulary includes the modality indicators and a series of semantic tokens indices in the form of $\langle\textit{AUD\_X}\rangle$, where $X$ ranges from 0 to 499, the same as the number of centroids of the audio tokenizer. In this design the prediction of semantic tokens is within LLM's vocabulary: After obtaining the indices which are in the form of such LLM-aware acoustic tokens, the semantic tokens are constructed by simply fetching the corresponding centroids. The extended audio-text vocabulary now becomes $V=\{V_t, V_a\}$. 

To further elaborate on the conditional generation tasks performed by LLM: for the VTA task, the source input $X_v = \{x_e^i\}_{i=1}^N$ is a sequence of embeddings and $x_e \in \mathbb{R}^D$, where $D$ is the embedding dimension of LLM. Our LLM backbone is a decoder-only structure with the next token prediction method. The distribution of the predicted token in the first layer is given by $p_{\theta_{LLM}}(\mathbf{C}_1 |X)=\prod_i p_{\theta_{LLM}}(c_1^i |X, \mathbf{C}_1^{<i})$ autoregressively. The objective has thus become: 
\begin{equation}
    \mathcal{L}_{\mathit{LLM}}=-\sum_{i=1}^{T'} \log p_{\theta_{\mathit{LLM}}}(c_1^i |X,\mathbf{C}_1^{<i}),
    \label{eq:llm:loss}
\end{equation}  
where $T'$ is the number of semantic tokens generated by LLM, $\theta_{\mathit{LLM}}$ is the parameter of LLM, $c_{1}^i$ is the token generated at step $i$, $\mathbf{C}_1^{<i}$ are previous tokens, and $X$ is the input condition. 

During inference, the LLM will autoregressively predict the next token until $\langle\textit{eos}\rangle$ is generated. Our LLM thus serves as the bridge for connecting between modalities. 

In our experiments, we use Gemma2-2B-it~\cite{team2024gemma}, a lightweight open-source LLM developed by Google, which is claimed to have comparable performance to a much larger variant Gemma-2-9B. We use Low-Rank Adaptor (LoRA)~\cite{hu2021lora} to finetune Gemma to make it understand vision/text conditions and generate audio tokens. 

\begin{figure*}
    \centering
    \includegraphics[height=0.3\textheight]{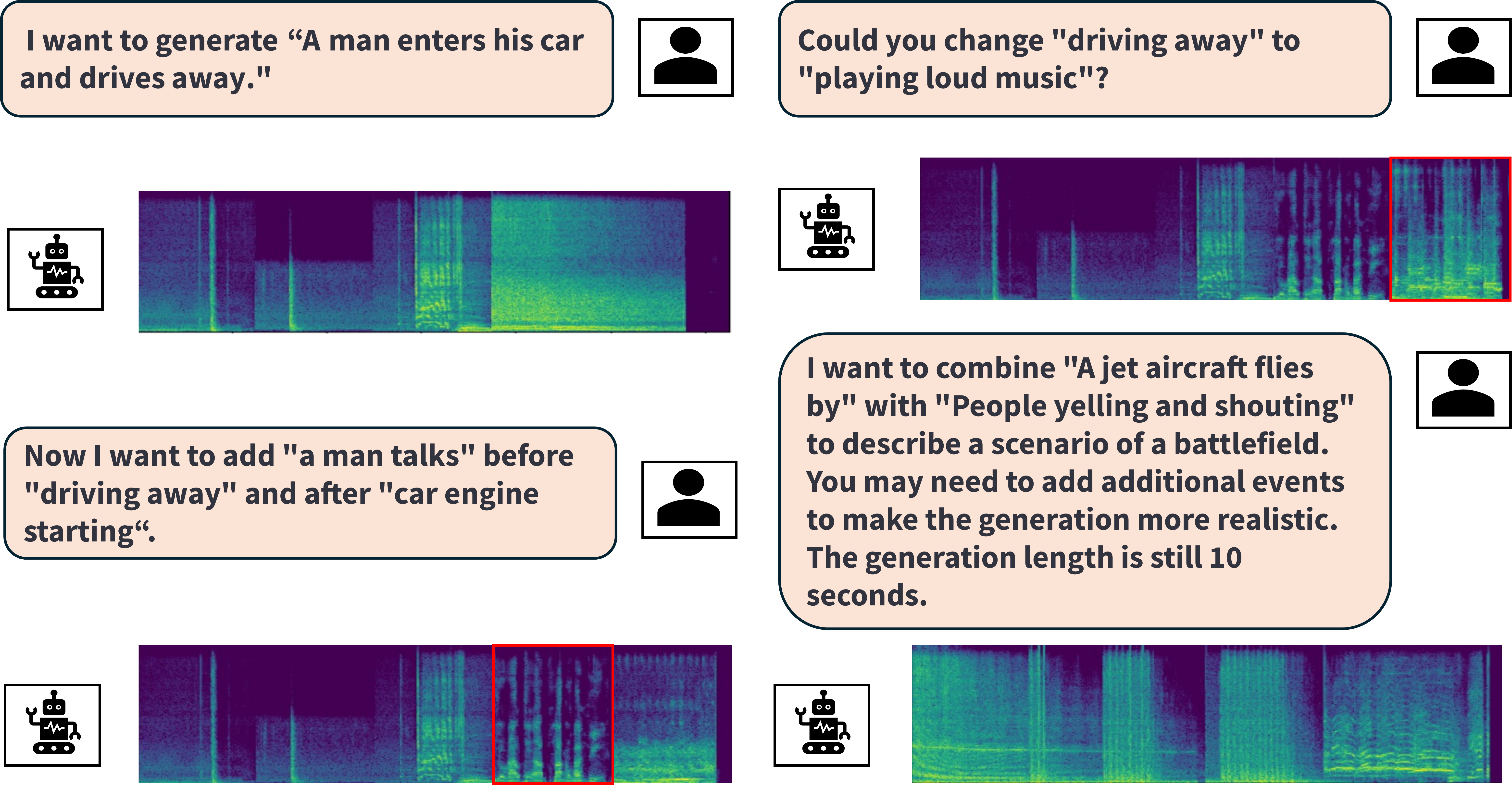}
    \vspace{-0.1in}
    \caption{A demo example showing Audio-Agent's conversation ability: First turn: Audio Generation; second turn: Audio Insertion; third Turn: Audio Editing; last turn: Audio Composition with high-level semantic instructions.
    Audio-Agent can choose to respond based on previous turns or make independent generations. {\it The corresponding audio files can be found in the supplementary materials.}}
    \label{fig.conversation_demo}
    \vspace{-0.15in}
\end{figure*}

\subsection{Conditional Audio Generation}
The audio generation module contains a diffusion model, text-based cross-attention layers and visual-based cross-attention layers. See Figure~\ref{fig.arch-backbone}. Given a query feature $Z$, text features $c_{txt}$ and visual features $c_{vis}$ the output for combining two types of cross-attention is defined as follows:
\begin{equation}
\begin{split}
\!\!\!\!\!\!\mathbf{Z}^{new}=\text{Softmax}(\frac{\mathbf{Q}\mathbf{K}_{txt}^{\top}}{\sqrt{d}})\mathbf{V}_{txt}+\text{Softmax}(\frac{\mathbf{Q}(\mathbf{K}_{vis})^{\top}}{\sqrt{d}})\mathbf{V}_{vis}\\
 \noindent\text{where}\ \mathbf{Q}=\mathbf{Z}\mathbf{W}_{txt}^q,\mathbf{K}_{txt}=\boldsymbol{c}_{txt}\mathbf{W}_{txt}^k, \mathbf{V}_{txt}=\boldsymbol{c}_{txt}\mathbf{W}_{txt}^v, \\
\mathbf{K}_{vis}=\boldsymbol{c}_{vis}\mathbf{W}_{vis}^k, \mathbf{V}_{vis}=\boldsymbol{c}_{vis}\mathbf{W}_{vis}^v
\end{split}
\end{equation}
The diffusion model and text-based cross-attention layers are from the pre-trained Auffusion model. During training, we keep the pre-trained part frozen. For the TTA task, we directly feed the step instructions as text conditions and arrange the output based on the start time and end time, as illustrated in Section~\ref{GPT}. For the VTA task, after obtaining the semantic tokens, we fetch the centroids from the HuBERT model according to the value indices as visual features. Similar to the text-based condition mechanism, we apply cross-attention on layers of the diffusion model. During inference, we introduce another parameter for controlling text and visual guidance:
\begin{equation}
\!\!\!\!\mathbf{Z}^{new}=\text{Attention}(\mathbf{Q},\mathbf{K}_{txt},\mathbf{V}_{txt}) + \lambda\cdot\text{Attention}(\mathbf{Q},\mathbf{K}_{vis},\mathbf{V}_{vis})
\end{equation}

Thus the final objective for the diffusion process, which is similar to latent diffusion models, is
\begin{equation}
L_{\text{simple}}=\mathbb{E}_{\boldsymbol{x}_{0},\boldsymbol{\epsilon}, \boldsymbol{c}_{txt}, \boldsymbol{c}_{vis}, t} \| \boldsymbol{\epsilon}- \boldsymbol{\epsilon}_\theta\big(\boldsymbol{x}_t, \boldsymbol{c}_{txt}, \boldsymbol{c}_{vis}, t\big)\|^2.
\end{equation}

Compared to IP-Adapter~\cite{ye2023ip}, our method introduces the video modality into audio generation. Furthermore, since the semantic tokens already incorporate temporal information of the video, we do not need to train an extra timestamp detection module as done by FoleyCrafter~\cite{zhang2024foleycrafter} to achieve temporal alignment. 

\subsection{Implementation Details}
For fine-tuning Gemma-2B-it, we set LoRA rank and alpha to be 64 with dropout to be 0.05. We separately train and fine-tune Gemma-2B-it, the projection layers and the cross-attention layers on the AVSync15~\cite{zhang2024audio} datasets. The training and evaluation are conducted on NVIDIA GeForce RTX 4090. Following~\cite{ye2023ip}, we set the $\lambda$ to be 0.5 as default.

%% file: sec/4_experiments.tex
\section{Experiments}
\subsection{Datasets}
Our framework is training-free for the TTA task. The evaluation of our complex generation ability is conducted on AudioCaps~\cite{kim2019audiocaps} dataset. We randomly choose either one caption from the test set or concatenate two of them together with the clause ``followed by." To achieve the optimal solution, the model needs to decompose the input into atomic text instructions, not only limited to two chunks. To better compare with other models, we limit our generation length to the standard 10 seconds. Following~\cite{xue2024auffusion}, we randomly selected 20 captions from each category for the generation. Additionally, to demonstrate Audio-Agent's ability to make inferences based on complex text conditions, we ask GPT to generate additional long event descriptions containing a series of events without explicitly describing the sound, such as ``A man enters his car and drives away". The number of complex captions is also 20. The baseline methods include AudioGen-v2-medium~\cite{kreuk2022audiogen}, AudioLDM2-large~\cite{liu2024audioldm}, Auffusion~\cite{xue2024auffusion}, Make-An-Audio 2~\cite{huang2023make} and WavJourney~\cite{liu2024wavjourney}.

We use AVSync15 for training and evaluation on VTA task. AVSync15 is a curated dataset from VGGSound Sync~\cite{chen2021audio} that has 1500 high video-audio alignment pairs, which is ideal for training and demonstrating temporal alignment between video and audio. Same experiment setting as~\cite{zhang2024foleycrafter} is used.
To better facilitate evaluation, we include some audio generation results in the supplementary material.

\begin{table}[ht]
\centering
\caption{Evaluation for all baseline models on the TTA task, categorized by the type of text conditions.}
\vspace{-0.1in}
\label{tab.tta_evaluation}
\scalebox{0.7}{
\begin{tabular}{@{}l|c|c|c}
\toprule
\multirow{2}{*}{Method}  & \textbf{Single Caption} & \textbf{Two Captions} & \textbf{Complex Captions}\\
& CLAP$\uparrow$ & CLAP$\uparrow$ &  CLAP$\uparrow$ \\ \midrule
AudioGen~\cite{kreuk2022audiogen}   & 49.34\% & 44.76\% & 23.98\%  \\
AudioLDM2~\cite{liu2024audioldm}      & 47.04\% & 36.03\% & 23.33\% \\
Auffusion~\cite{xue2024auffusion}          & 50.91\% & 45.90\% & 14.40\%  \\
Make-An-Audio 2~\cite{huang2023make} & ~\textbf{55.49}\% & 44.01\% & 23.73\% \\
WavJourney~\cite{liu2024wavjourney} & 34.83\% & 24.90\% & 23.71\%\\
Ours (with volume control)        & 55.26\% & 48.83\% & 21.10\%  \\
Ours                & 55.17\% & ~\textbf{53.02\%} & ~\textbf{24.06\%}  \\
\bottomrule
\end{tabular}
}
\vspace{-0.1in}
\end{table}

\begin{table}
\caption{Quantitative evaluation on semantic alignment and audio quality. Specifically, Audio-Agent achieves on par performance versus state-of-the-art models in terms of Mean KL Divergence (MKL)~\cite{iashin2021taming}, CLIP~\cite{wu2022wav2clip} and FID~\cite{heusel2017gans} on AVSync15 ~\cite{zhang2024audio}. {\em See text for explanation on the subpar performance.} } 
\vspace{-0.05in}
\label{tab.vta_result}
\centering
\resizebox{\linewidth}{!}{%
\begin{tabular}{lccc}
\toprule[1.5pt]
\multicolumn{1}{c}{Method} & \multicolumn{1}{c}{MKL  $\downarrow$} & \multicolumn{1}{c}{CLIP  $\uparrow$} & \multicolumn{1}{c}{FID  $\downarrow$}   \\ 
\midrule[1.5pt]
SpecVQGAN (Inception)~\cite{iashin2021taming}    &      5.339         &    6.610     &  114.44      \\
SpecVQGAN (ResNet)~\cite{iashin2021taming}        &3.603                  & 6.474    & 75.56\\
Diff-Foley~\cite{luo2024diff}             &    1.963                   & 10.38  &     65.77         \\
Seeing and Hearing~\cite{xing2024seeing}             &    2.547                      & 2.033  &      65.82                                          \\
FoleyCrafter~\cite{zhang2024foleycrafter}       & \textbf{1.497}        & \textbf{11.94}  &  \textbf{36.80} \\

Ours (without temporal connector)                 &    2.516                      &  9.06  &      55.59  \\

Ours (with temporal connector)                 &    2.623                      & 8.55  &      52.93  
\\\bottomrule[1.5pt]
\end{tabular}
}
\vspace{-0.15in}
\end{table}

\begin{table}
\caption{Quantitative evaluation in terms of temporal synchronization. We report onset detection accuracy (Onset ACC) and average precision (Onset AP) for the generated audios on AVSync~\cite{zhang2024audio}, which provides onset timestamp labels for assessment, following previous studies ~\cite{luo2024diff,xie2024sonicvisionlm}.}
\vspace{-0.05in}
\label{tab.vta_sync}
\centering
\resizebox{\linewidth}{!}{%
\begin{tabular}{lcc}
\toprule[1.5pt]
\multicolumn{1}{c}{Method} & \multicolumn{1}{c}{Onset ACC $\uparrow$} & \multicolumn{1}{c}{Onset AP  $\uparrow$}  \\ 
\midrule[1.5pt]
SpecVQGAN(Inception)~\cite{iashin2021taming}    & 16.81                              & 64.64                                      \\
SpecVQGAN(ResNet)~\cite{iashin2021taming}        & 26.74                  & 63.18                                     \\
Diff-Foley~\cite{luo2024diff}             & 21.18                       & 66.55                       \\
Seeing and Hearing~\cite{xing2024seeing}             &    20.95                      & 60.33                         \\
FoleyCrafter~\cite{zhang2024foleycrafter}        & 28.48                           &  68.14                                   \\
Ours (without temporal connector)        & 28.45                   & 64.72        \\
Ours (with temporal connector)        &  \textbf{29.01}                 & \textbf{69.38}  \\
\bottomrule[1.5pt]
\end{tabular}
}
\vspace{-0.15in}
\end{table}

\subsection{Evaluation Metrics}
The evaluation metrics are summarized as follows: For the VTA task, we use the Frechet audio distance (FAD) to evaluate audio fidelity. Additionally, we utilize the MKL metric~\cite{iashin2021taming} and CLIP similarity~\cite{wu2022wav2clip} for audio-video relevance. Furthermore, to evaluate the synchronization of the generated audio in the video-to-audio setting, we use the same evaluation metrics as CondFoleyGen~\cite{du2023conditional}, namely \# Onset Accuracy, and Onset AP. For the TTA task, we use CLAP similarity~\cite{wu2023large} 
\subsection{Evaluation and comparison}
\begin{figure}
    \centering
    \includegraphics[width=\linewidth]{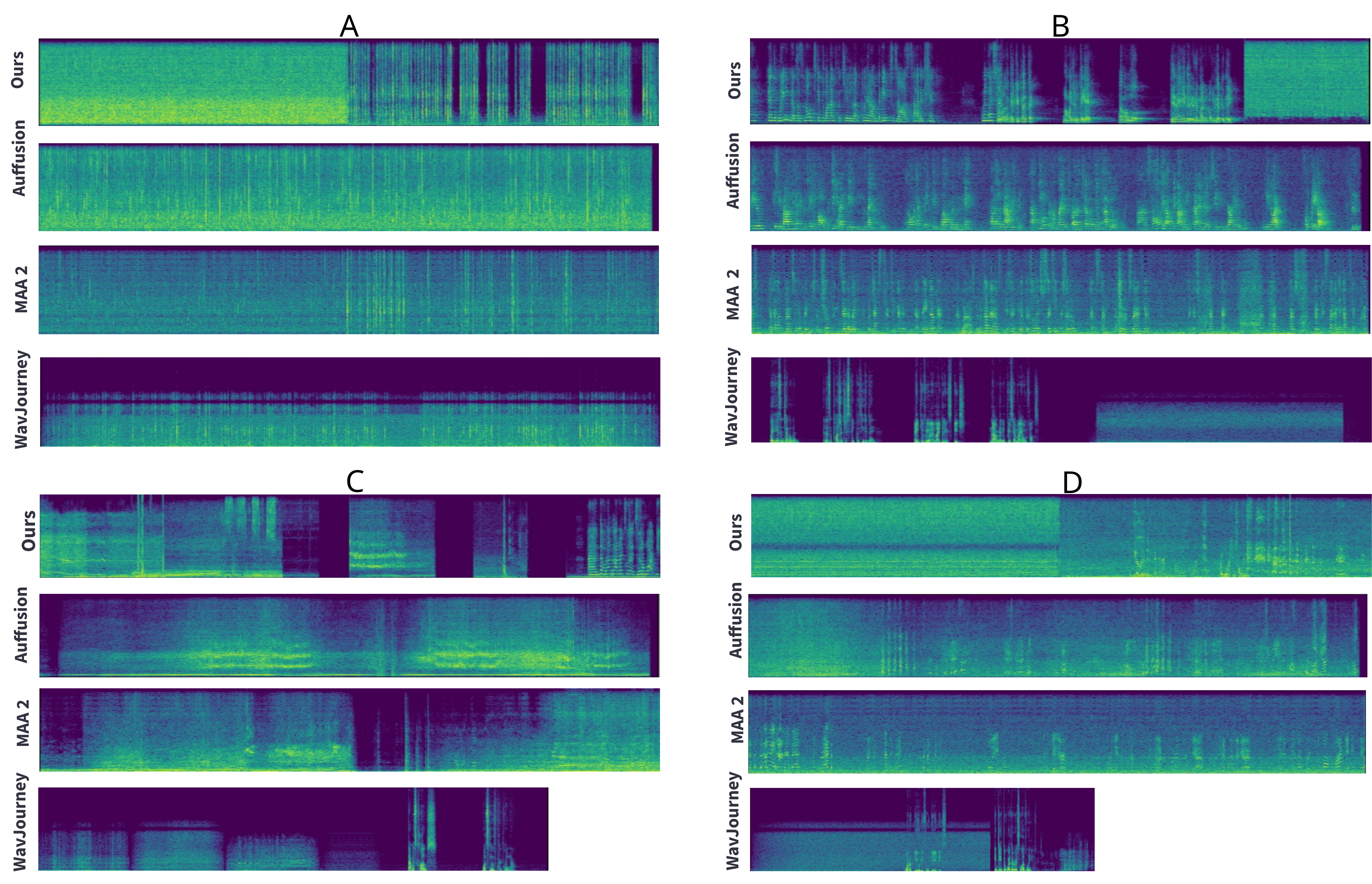}
    \vspace{-0.2in}
    \caption{Comparison with baseline on TTA. From top: our Audio-Agent, Auffusion, Make-An-Audio 2, and WavJourney. To demonstrate audio generation based on long complex text conditions, we ask the respective model to generate audio clips for 20 seconds. The text condition is drawn from the Two Captions category of Table~\ref{tab.tta_evaluation}: (A)~A river stream of water flowing followed by typing on a computer keyboard; (B)~A woman delivering a speech followed by a male speech and statics; (C)~A vehicle engine revving then accelerating at a high rate as a metal surface is whipped followed by tires skidding followed by a door shutting and a female speaking; (D)~ Continuous white noise followed by a vehicle driving as a man and woman are talking and laughing. Observe (by {\it listening to the corresponding audio files in the supplementary material}) our method successfully generates multi-event audio at different times based on descriptions, while other baselines mix the generated audio. Particularly for WavJourney, although the boundary between events is more distinct, it consistently does not obey the prompt to generate 20 seconds of audio. The generated audio clips are respectively of length 20 seconds, 30 seconds (truncated here), 16 seconds, and 12 seconds.}
    \label{fig.tta-example}
    \vspace{-0.15in}
\end{figure}

Audio-Agent outperforms other baseline methods on the majority of TTA experiment settings, only lagging behind Make-an-Audio 2 by a small margin on the Single Caption setting, which we attribute to the factor of our generation backbone, since we did not train our diffusion model on complex caption datasets. The result can be found in Table~\ref{tab.tta_evaluation}. Moreover, our method outperforms the original Auffusion backbone model by a significantly increasing margin as the text condition becomes longer and more complex. From the visualization examples in Table~\ref{fig.tta-example}, we can see our method achieves more fine-grained distinction among events with better control on generation length. Specifically, we notice that with a longer text condition, all of AudioGen~\cite{kreuk2022audiogen}, AudioLDM2~\cite{liu2024audioldm}, Auffusion~\cite{xue2024auffusion} and Make-an-Audio 2~\cite{huang2023make} exhibit miss-out events. For example, if the text condition is multi-event such as ``\underline{Pigeons cooing and bird wings flapping} as \underline{footsteps shuffle on paper} followed by \underline{motor sounds with male speaking}", all the above baseline methods fail to generate the motor sound at the end of the audio clip during evaluation. In addition, during our experiments, we found that WavJourney often struggles to generate audio with the precise duration specified by the user, even when the duration is explicitly included in the prompt. Our method avoids these issues by using GPT-4 as a coordinator for effective caption analysis and generation planning, enabling finer distinctions between events. In addition, we found that if the text description contains too many events, the GPT-4 tends to lower the volume of every event to prevent possible volume explosion. Although this improves the sound quality from the human ear, the distinction between each event becomes less clear, which may result in a decrease in CLAP score.

We also notice a significant drop for all these methods on complex captions, since none of them has been trained on this type of text condition which is closer to real-world situations,
%Still, we find this type of text condition more practical in the real world,
since it does not require meticulous and explicit descriptions of the characteristics of the sound, but rather describes the scenario for sound generation, thus offering more flexibility for the sound producer. In the supplementary materials, we provide examples of complex results used for evaluation, along with additional comparisons of multi-turn generation and editing.

\begin{figure}
    \centering
    \includegraphics[width=\linewidth]{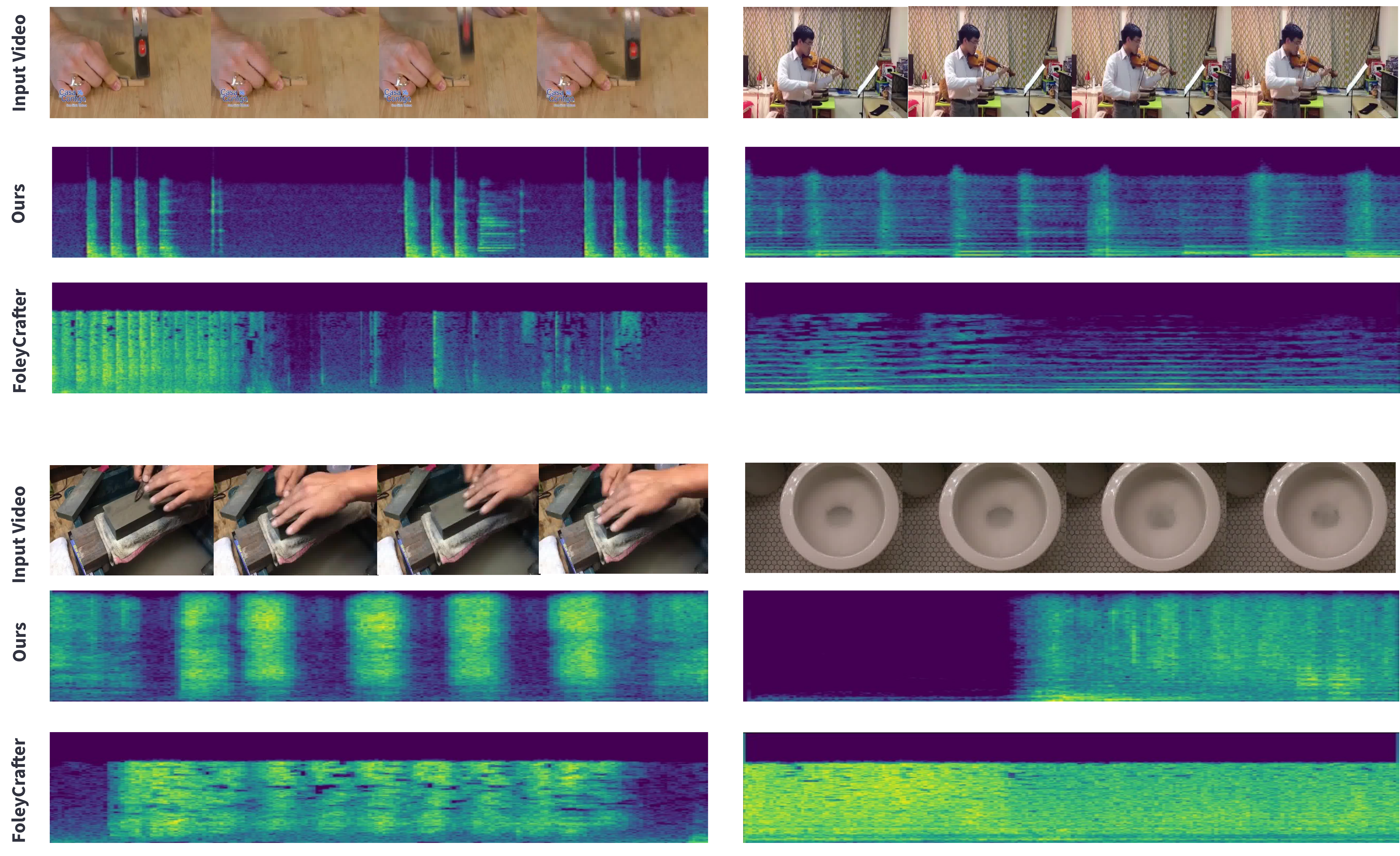}
    \vspace{-0.2in}
    \caption{Comparison with baseline for VTA generation task. Compared to the baseline, the event occurrence is more explicit. Our method can produce audio that is more aligned and better synchronized with the input video. {\it The corresponding audio files can be found in the supplementary materials.}}
    \label{fig.sync}
    \vspace{-0.15in}
\end{figure}

For the VTA task, our method achieves better visual-audio synchronization compared to other baseline methods, while subpar compared with the current state-of-the-art method in terms of generation audio quality, see Tables~\ref{tab.vta_result} and~\ref{tab.vta_sync}. This is indeed expected, 
%We consider this reasonable 
as most of the other baseline methods have been trained on multiple larger datasets.  
We also find that the temporal connector may negatively affect the generated audio quality on a small scale. However, for the evaluation of synchronization, we noticed a significant improvement after the temporal connector was applied, especially for the Onset AP. Without explicit training of a timestamp detector, our method achieves a better performance across all baseline methods in terms of onset Acc and Onset AP, see Figure~\ref{fig.sync} for illustration.  

\subsection{Ablation Studies}
We include our ablation study on different LoRA rank values during LLM fine-tuning, see Tables~\ref{tab.ablation_vta_quality} and~\ref{tab.ablation_vta_sync}. We find that an increase in trainable parameters sometimes does not necessarily improve the result. Notwithstanding, for a fair comparison, we use the rank value of 60 across all metrics. Additionally during training, we find that the training of the cross-attention layer can converge within 20,000 steps. We notice that the loss curve is not a reliable indicator of the model's performance. The model can achieve a good performance even when the loss curve remains flat.

Additionally, our framework can be easily integrated with different LLMs for action planning and conversation, apart from GPT-4. In the supplementary material, we included the comparison between various models such as Llama~\cite{touvron2023llama}, Gemma~\cite{team2024gemma}, and Mixtral~\cite{jiang2024mixtral}. With this, our framework's performance can be improved with the development of LLMs in more precise instruction-following with more sophisticated prompts.
\begin{table}
\caption{Ablation study on AVSync15 dataset with different LoRA rank for semantic alignment and audio quality. During experiments, we keep the value of alpha the same as the rank. {\em See text for further explanation.}} 
\vspace{-0.05in}
\label{tab.ablation_vta_quality}
\centering
\resizebox{\linewidth}{!}{%
\begin{tabular}{lcccc}
\toprule[1.5pt]

\multicolumn{1}{c}{Method} & \multicolumn{1}{c}{Trainable Parameters} & \multicolumn{1}{c}{MKL  $\downarrow$} & \multicolumn{1}{c}{CLIP  $\uparrow$} & \multicolumn{1}{c}{FID  $\downarrow$}   \\ 
\midrule[1.5pt]
Ours (R=16)    & 78.31MM             &    2.702                      & 8.42  &      58.426  \\
Ours (R=32)    &99.08MM             &    ~\textbf{2.543}                      & 8.49  &      55.197  \\
Ours (R=64)    &140.61MM             &    2.623                      & ~\textbf{8.55}  &      ~\textbf{52.929}  \\
\bottomrule[1.5pt]
\end{tabular}
}
\vspace{-0.15in}
\end{table}

\begin{table}
\caption{Ablation study on AVSync15 dataset with different LoRA rank in terms of temporal synchronization. During experiments, we keep the value of alpha the same as the rank. {\em See text for further explanation.}}
\vspace{-0.05in}
\label{tab.ablation_vta_sync}
\centering
\resizebox{\linewidth}{!}{%
\begin{tabular}{lccc}
\toprule[1.5pt]
\multicolumn{1}{c}{Method} & \multicolumn{1}{c}{Trainable Parameters} & \multicolumn{1}{c}{Onset ACC $\uparrow$} & \multicolumn{1}{c}{Onset AP  $\uparrow$}  \\ 
\midrule[1.5pt]
Ours (R=16)     &78.31M   & ~\textbf{29.74}                   & ~\textbf{70.63}        \\
Ours (R=32)     &99.08M   & 27.49                  & 70.57       \\
Ours (R=64)     &140.61M   & 29.01                   & 69.38        \\
\bottomrule[1.5pt]
\end{tabular}
}
\vspace{-0.15in}
\end{table}

%% file: sec/5_conclusion.tex
\section{Conclusion and Discussion}
%\subsection{Limitation and Future Work}
\noindent\textbf{Limitation and Future Work} Our framework experiences a performance drop when text conditions becomes increasingly complex for the TTA task, where the drop is more severe in other baseline methods. We believe it is a worthwhile direction in the future for understanding long and complex captions with improved fine-grained delineations among multiple and overlapping events. We may also utilize the LLM's versatility involving audio captioning tasks and video captioning tasks. The above are worthwhile future directions to explore.

%\subsection{Conclusion}
\vspace{2mm}
\noindent\textbf{Conclusion}
In this paper, we present Audio-Agent, a multimodal framework for both text-to-audio and video-to-audio tasks. Our model contributes a subbstantial first attempt in offering a conversation-based method for audio generation, editing and composition, facilitating audio generation conditioned on multievent complex descriptions. For the video-to-audio task, we propose an efficient method to achieve visual synchronization. Through extensive experiments, we show that our model can synthesize high-fidelity audio, ensuring high semantic alignment with input. Moreover, our work represents an initial and significant step toward multi-event long-condition TTA generation which has not been fully explored.

%% file: sec/X_suppl.tex
\clearpage
\setcounter{page}{1}
\maketitlesupplementary

\section{supplementary material}
\subsection{Prompt example for TTA task} 
We provide our prompt instruction in Table~\ref{tab.prompt_instruction_tta}, the volume control variation in Table~\ref{tab.prompt_instruction_tta_volume}, and in context examples in Tables~\ref{tab.in_context_examples_tta}, ~\ref{tab.in_context_examples_tta_continue},~\ref{tab.in_context_examples_tta_volume} and ~\ref{tab.in_context_examples_tta_volume_continue}.

\begin{table*}[p]
\setlength\tabcolsep{0pt}
\centering
\caption{Our prompt instruction for TTA generation}
\label{tab.prompt_instruction_tta}
\begin{tabular*}{\linewidth}{@{\extracolsep{\fill}} l }\toprule
\begin{lstlisting}[style=myverbatim]
**You are a dialog agent that assists users in generating audio through conversation. The user begins by describing the audio they envision, and you help translate this description into multiple audio captions suitable for generating. You have a powerful tool at your disposal, Auffusion, which can generate simple, atomic audio based on textual descriptions. Your task is to determine how best to utilize this tool, which may involve multiple calls to Auffusion to produce a complex audio sequence composed of simpler audio.**

**Here are 10 examples of the types of descriptions Auffusion was trained on. These should guide you in understanding what constitutes a “simple” and “atomic” motion:**
1. A muddled noise of broken channel of the TV.
2. A person is turning a map over and over.
3. Several barnyard animals mooing in a barn.
4. An office chair is squeaking.
5. A flying bee is buzzing loudly around an object.
6. Thunder claps far in the distance.
7. Something goes round that is playing its song.
8. A paper printer is printing off multiple pages.
9. A person is making noise by tapping their fingernails on a solid surface.
10.A person crunches through dried leaves on the ground.

**Instructions:**
1. **User-Provided Description**: The user's description will include both straightforward and complex descriptions of audio. The user may also provide multiple descriptions and ask you to combine them together.
2. **Auffusion Invocation**: For each audio description, you must decide how to break down the description into simple, atomic audio. Invoke the Auffusion API to generate each component of the audio sequence. Ensure that each call focuses on a straightforward, non-elaborate audio description.
3. **Plan Generation**: Your response should include a step-by-step plan detailing each call to Auffusion necessary to create the complete audio sequence. 
4. **Requirement**: 
4.1. You should include the start_time and end_time in this call. The audio length is 10 seconds, and thus you should have at least one call having end_time=10. 
4.2. If the user input has multiple events or asks to combine multiple description together, you should have overlapping audios happening in the same range of time. There should have less than three audios in the same time. Overlapping means one audio having smaller start_time than another audio's end_time
4.3. You're free to generate as many as calls you like, but please keep the minimum number of calls. 

**Response Format:**
- You should only respond in JSON format, following this template:
```json
{
  "plan": "A numbered list of steps to take that conveys the long-term plan"
}
```
\end{lstlisting} \\\bottomrule
\end{tabular*}
\end{table*}

\begin{table*}[p]
\setlength\tabcolsep{0pt}
\centering
\caption{Our in-context examples for TTA generation.}
\label{tab.in_context_examples_tta}
\begin{tabular*}{\linewidth}{@{\extracolsep{\fill}} l }\toprule
\begin{lstlisting}[style=myverbatim]
**Examples:**

**Example 1:**
- **User Input**: I want to generate "A clap of thunder coupled with the running water".
- **Your Output**:
```json
{
  "plan": "1. Auffusion.generate('A clap of thunders.',start_time=2,end_time=5); 2. Auffusion.generate('Rain pouring outside.',start_time=0, end_time=10)"
}
```

**Example 2:**
- **User Input**: I want to combine "Buzzing and humming of a motor" with "A man speaking" together
- **Your Output**:
```json
{
  "plan": "1. Auffusion.generate('A motor buzzing and humming',start_time=0,end_time=10); 2. Auffusion.generate('A man speaking.',start_time=3,end_time=6)"
}
```

**Example 3:**
- **User Input**: I want to generate "A series of machine gunfire and two gunshots firing as a jet aircraft flies by followed by soft music playing"
- **Your Output**:
```json
{
  "plan": "1. Auffusion.generate('A series of machine gunfire.',start_time=0,end_time=4); 2. Auffusion.generate('Two gunshots firing.',start_time=4,end_time=6); 3. Auffusion.generate('A jet aircraft flies.',start_time=0,end_time=6); 4. Auffusion.generate('Soft music playing.',start_time=6,end_time=10)"
}
```
\end{lstlisting} \\\bottomrule
\end{tabular*}
\end{table*}

\begin{table*}[p]
\setlength\tabcolsep{0pt}
\centering
\caption{Our in-context examples for TTA generation (continue).}
\label{tab.in_context_examples_tta_continue}
\begin{tabular*}{\linewidth}{@{\extracolsep{\fill}} l }\toprule
\begin{lstlisting}[style=myverbatim]
**Example 4:**
- **User Input**: I want to generate "A crowd of people playing basketball game."
- **Your Output**:
```json
{
  "plan": "1. Auffusion.generate('Sound of a basketball bouncing on the court.',start_time=0, end_time=7); 2. Auffusion.generate('A ball hit the basket',start_time=5, end_time=7); 3. Auffusion.generate('People cheering and shouting.',start_time=7, end_time=10)"
}
```
- **Followed up User Input**: I want to change it to "people playing table tennis".
- **Your Output**:
```json
{ 
  "plan": "1. Auffusion.generate('Sound of a table tennis ball bouncing on the table.',start_time=0,end_time=7); 2. Auffusion.generate('People cheering and shouting.',start_time=7,end_time=10)" 
}
```
```
\end{lstlisting} \\\bottomrule
\end{tabular*}
\end{table*}

\begin{table*}[p]
\setlength\tabcolsep{0pt}
\centering
\caption{Our prompt instruction for TTA generation}
\label{tab.prompt_instruction_tta_volume}
\begin{tabular*}{\linewidth}{@{\extracolsep{\fill}} l }\toprule
\begin{lstlisting}[style=myverbatim]
**You are a dialog agent that assists users in generating audio through conversation. The user begins by describing the audio they envision, and you help translate this description into multiple audio captions suitable for generating. You have a powerful tool at your disposal, Auffusion, which can generate simple, atomic audio based on textual descriptions. Your task is to determine how best to utilize this tool, which may involve multiple calls to Auffusion to produce a complex audio sequence composed of simpler audio.**

**Here are 10 examples of the types of descriptions Auffusion was trained on. These should guide you in understanding what constitutes a “simple” and “atomic” motion:**
1. A muddled noise of broken channel of the TV.
2. A person is turning a map over and over.
3. Several barnyard animals mooing in a barn.
4. An office chair is squeaking.
5. A flying bee is buzzing loudly around an object.
6. Thunder claps far in the distance.
7. Something goes round that is playing its song.
8. A paper printer is printing off multiple pages.
9. A person is making noise by tapping their fingernails on a solid surface.
10.A person crunches through dried leaves on the ground.

**Instructions:**
1. **User-Provided Description**: The user’s description will include both straightforward and complex descriptions of audio. The user may also provide multiple descriptions and ask you to combine them together.
2. **Auffusion Invocation**: For each audio description, you must decide how to break down the description into simple, atomic audio. Invoke the Auffusion API to generate each component of the audio sequence. Ensure that each call focuses on a straightforward, non-elaborate audio description.
3. **Plan Generation**: Your response should include a step-by-step plan detailing each call to Auffusion necessary to create the complete audio sequence. 
4. **Requirement**: 
4.1. You should include the start_time and end_time in this call. The audio length is 10 seconds, and thus you should have at least one call having end_time=10. 
4.2. You should include the volume for each generation call in dB following LUFS standard.
4.3. If the user input has multiple events or asks to combine multiple description together, you should have overlapping audios happening in the same range of time. There should have less than three audios in the same time. Overlapping means one audio having smaller start_time than another audio's end_time. You should correspondingly adjust the volume for each overlapping calls.
4.4. You're free to generate as many as calls you like, but please keep the minimum number of calls. 

**Response Format:**
- You should only respond in JSON format, following this template:
```json
{
  "plan": "A numbered list of steps to take that conveys the long-term plan"
}
```
\end{lstlisting} \\\bottomrule
\end{tabular*}
\end{table*}

\begin{table*}[p]
\setlength\tabcolsep{0pt}
\centering
\caption{Our in-context examples for TTA generation.}
\label{tab.in_context_examples_tta_volume}
\begin{tabular*}{\linewidth}{@{\extracolsep{\fill}} l }\toprule
\begin{lstlisting}[style=myverbatim]
**Examples:**

**Example 1:**
- **User Input**: I want to generate "A clap of thunder coupled with the running water".
- **Your Output**:
```json
{
  "plan": "1. Auffusion.generate('A clap of thunders.',start_time=2,end_time=5,volume=-10); 2. Auffusion.generate('Rain pouring outside.',start_time=0, end_time=10,volume=-30)"
}
```

**Example 2:**
- **User Input**: I want to combine "Buzzing and humming of a motor" with "A man speaking" together
- **Your Output**:
```json
{
  "plan": "1. Auffusion.generate('A motor buzzing and humming',start_time=0,end_time=10,volume=-25); 2. Auffusion.generate('A man speaking.',start_time=3,end_time=6,volume=-15)"
}
```

**Example 3:**
- **User Input**: I want to generate "A series of machine gunfire and two gunshots firing as a jet aircraft flies by followed by soft music playing"
- **Your Output**:
```json
{
  "plan": "1. Auffusion.generate('A series of machine gunfire.',start_time=0,end_time=4,volume=-10); 2. Auffusion.generate('Two gunshots firing.',start_time=4,end_time=6,volume=-10); 3. Auffusion.generate('A jet aircraft flies.',start_time=0,end_time=6,volume=-15); 4. Auffusion.generate('Soft music playing.',start_time=6,end_time=10,volume=-20)"
}
```
\end{lstlisting} \\\bottomrule
\end{tabular*}
\end{table*}

\begin{table*}[p]
\setlength\tabcolsep{0pt}
\centering
\caption{Our in-context examples for TTA generation (continue).}
\label{tab.in_context_examples_tta_volume_continue}
\begin{tabular*}{\linewidth}{@{\extracolsep{\fill}} l }\toprule
\begin{lstlisting}[style=myverbatim]
**Example 4:**
- **User Input**: I want to generate "A crowd of people playing basketball game."
- **Your Output**:
```json
{
  "plan": "1. Auffusion.generate('Sound of a basketball bouncing on the court.',start_time=0, end_time=7,volume=-15); 2. Auffusion.generate('A ball hit the basket',start_time=5, end_time=7,volume=-10); 3. Auffusion.generate('People cheering and shouting.',start_time=7, end_time=10,volume=-10)"
}
```
- **Followed up User Input**: I want to change it to "people playing table tennis".
- **Your Output**:
```json
{ 
  "plan": "1. Auffusion.generate('Sound of a table tennis ball bouncing on the table.',start_time=0,end_time=7,volume=-10); 2. Auffusion.generate('People cheering and shouting.',start_time=7,end_time=10,volume=-10)" 
}
```
\end{lstlisting} \\\bottomrule
\end{tabular*}
\end{table*}

\subsection{Prompt example for VTA task} 
We provide our prompt instruction in Table~\ref{tab.prompt_instruction_vta}. The prompt format follows the requirement from Gemma2-2B-it.

\begin{table*}[p]
\setlength\tabcolsep{0pt}
\centering
\caption{Our prompt instruction for VTA generation}
\label{tab.prompt_instruction_vta}
\begin{tabular*}{\linewidth}{@{\extracolsep{\fill}} l }\toprule
\begin{lstlisting}[style=myverbatim]
<start_of_turn>user
You are an intelligent audio generator for videos.
You don’t need to generate the videos themselves but need to generate the audio suitable for the video, with sementic coherence and temporal alignment.
I'll give you the video embedding enclosed by <Video></Video>, also the video caption enclosed by <Caption></Caption>.
Your goal is to generate the audio indices for the video
You only need to output audio indices, such as <AUD_x>, where x is the index number.

Your turn:
Given the video <Video><VideoHere></Video> and the video caption <Caption><CaptionHere></Caption>, the accompanied audio for the video is: 

<end_of_turn>
<start_of_turn>model

\end{lstlisting} \\\bottomrule
\end{tabular*}
\end{table*}

\subsection{Complex captions for TTA task} 
We provide examples of GPT-generated complex captions in Table~\ref{tab.complex_caption_tta_eval} that we use for TTA task evaluation.

\subsection{Comparison with different LLM for TTA task}
We present results using different LLMs for generation planning in
Table~\ref{tab:llm_ablation}. We found that most LLMs can follow the instruction and output atomic descriptions, except Gemma2-27B and Llama-3.1-8B output contains repetitive or unnecessary actions. With stronger backbone, our framework could achieve improved results.

\begin{table*}[p]
\setlength\tabcolsep{0pt}
\centering
\caption{Examples of our complex caption for TTA generation}
\label{tab.complex_caption_tta_eval}
\begin{tabular*}{\linewidth}{@{\extracolsep{\fill}} l }\toprule
\begin{lstlisting}[style=myverbatim]
1. A man enters his car and drives away
2. A couple decorates a room, hangs pictures, and admires their work.
3. A mechanic inspects a car, changes the oil, and test drives the vehicle.
4. A group of kids play hide and seek in a large, old house.
5. A woman packs a suitcase, locks her house, and walks to the bus station.
\end{lstlisting} \\\bottomrule
\end{tabular*}
\end{table*}

\begin{table}[!ht]
\centering
\renewcommand{\arraystretch}{1.5}
\begin{tabular}{p{2cm}p{10cm}} % Increased from 9cm to 10cm
\hline
\textbf{LLM} & \textbf{Turn 1 Decomposition Results}\\ \hline
GPT-4-Turbo         & \parbox[t]{10cm}{\raggedright 
1. 'Car door opening and closing.', start\_time=0, end\_time=3; \\
2. 'Car engine starts.', start\_time=2, end\_time=5 \\
3. 'Car driving away.', start\_time=4, end\_time=10} \\
\hline
Gemma2-27B          &\parbox[t]{10cm}{\raggedright
1. 'A car door opening and closing.', start\_time=0, end\_time=2\\
2. 'A man closing car door.', start\_time=2, end\_time=4; \\
3. 'A car engine starting.', start\_time=4, end\_time=5\\
4. 'A car driving away.', start\_time=5, end\_time=10} \\
\hline
Llama-3.1-8B      & \parbox[t]{10cm}{\raggedright 
1. 'A car engine starts and the car drives away.',start\_time=0,end\_time=5 \\
2. 'A man enters his car and drives away.',start\_time=0,end\_time=5 \\ 
3. 'A man is talking on the phone.',start\_time=5,end\_time=10} \\
\hline
Llama-2-7B & \parbox[t]{10cm}{\raggedright 
1. 'A car door opens.',start\_time=0,end\_time=1\\
2. 'A man gets in the car.',start\_time=1,end\_time=3 \\ 
3. 'Car engine starts.',start\_time=3,end\_time=5 \\
4. 'Car drives away.',start\_time=5,end\_time=10} \\
\hline
Mixtral-8x7B-v0.1 & \parbox[t]{10cm}{\raggedright 
1. 'Car door opening.',start\_time=0, end\_time=3 \\
2. 'A man getting into the car.',start\_time=1, end\_time=4 \\
3. 'Car door closing.',start\_time=3, end\_time=5 \\
4. 'Car engine starting.',start\_time=4, end\_time=6 \\ 5. 'Car driving away.',start\_time=6, end\_time=10
}\\
\hline
Meta-Llama-3-70B & \parbox[t]{10cm}{\raggedright
1. 'A car door opens and closes.',start\_time=0,end\_time=2 \\
2. 'A person gets in a car and settles in.',start\_time=0,end\_time=2 \\
3. 'A car engine starts.',start\_time=2,end\_time=4\\
4. 'A car drives away.',start\_time=4,end\_time=10} \\
\hline
\end{tabular}
\caption{Comparison of Using Different LLMs on the caption "A man enters his car and drives away.".}
\label{tab:llm_ablation}
\end{table}